\documentclass{article}

\usepackage[pdftex]{graphicx}

\usepackage[normalem]{ulem}
\usepackage{siunitx}
\usepackage{color}

\usepackage[cmex10]{amsmath}

\begin{document}

\title{Phase asymmetry effect in longitudinal offset coupled resonator optical waveguides}

\author{Pedro~Chamorro-Posada and  F. Javier Fraile-Pel\'aez}

\maketitle
.
\begin{abstract}
We show that the implementation of the longitudinal displacement technique for adjusting the coupling coefficients in microring waveguides is subject to a phase asymmetry effect.  This issue is shown to substantially alter the system response in apodized filters and cannot be ignored in the design stage.   
\end{abstract}

\section{Introduction}

\begin{figure}[!t]
\centering
\begin{tabular}{c}
(a)\\
\includegraphics[width=2in]{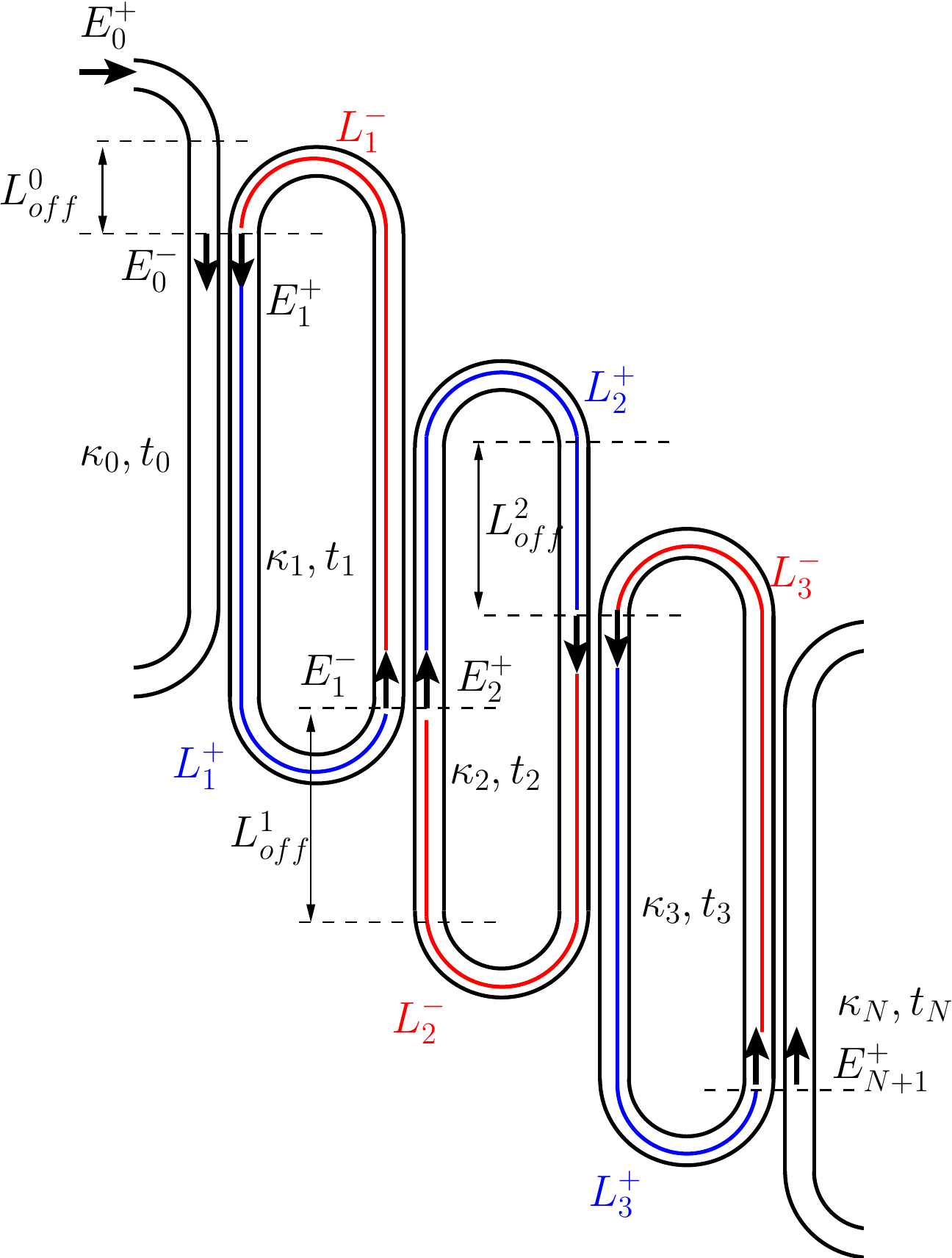}\\
(b)\\
\includegraphics[width=2in]{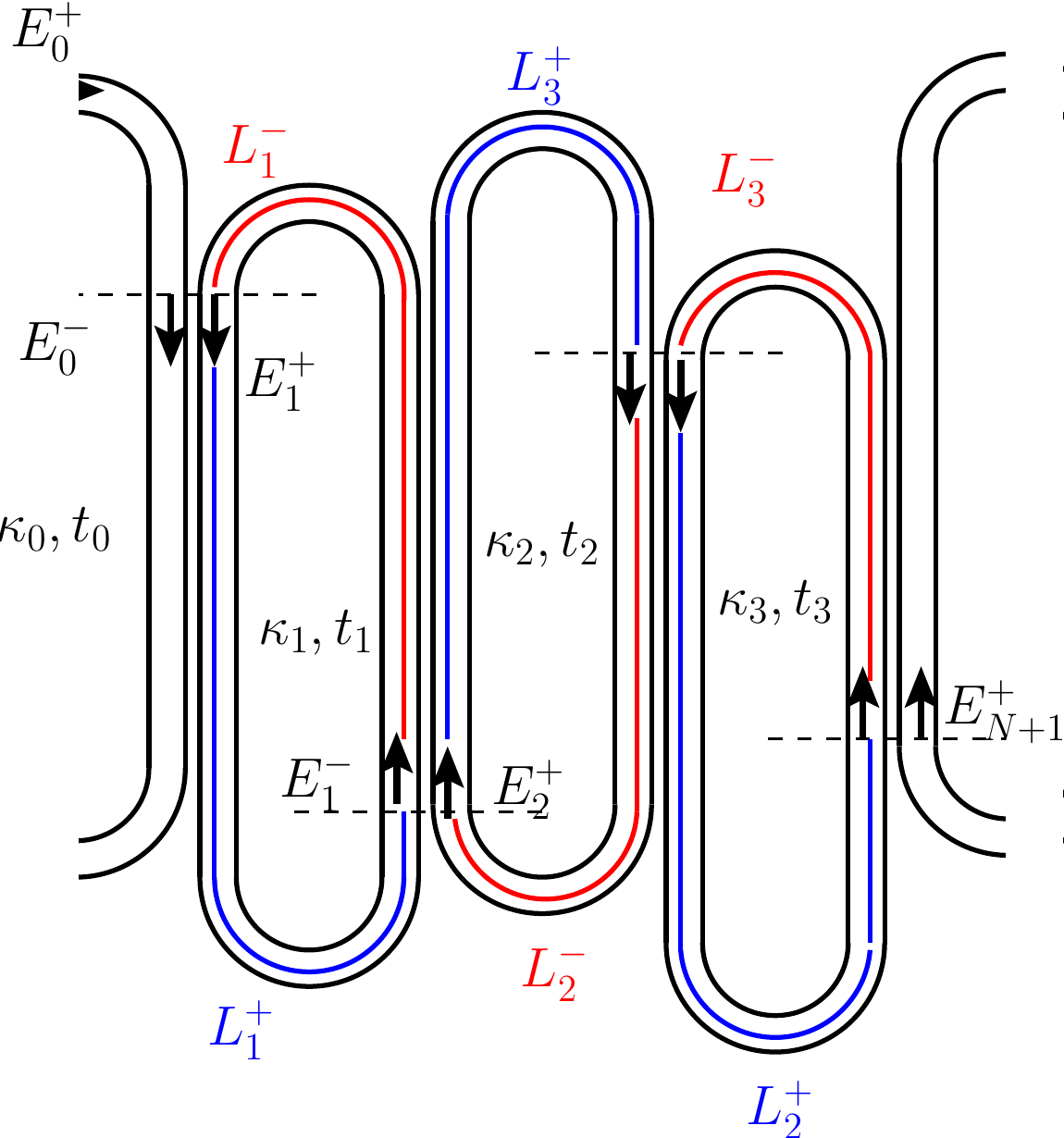}\\
(c)\\
\includegraphics[width=2in]{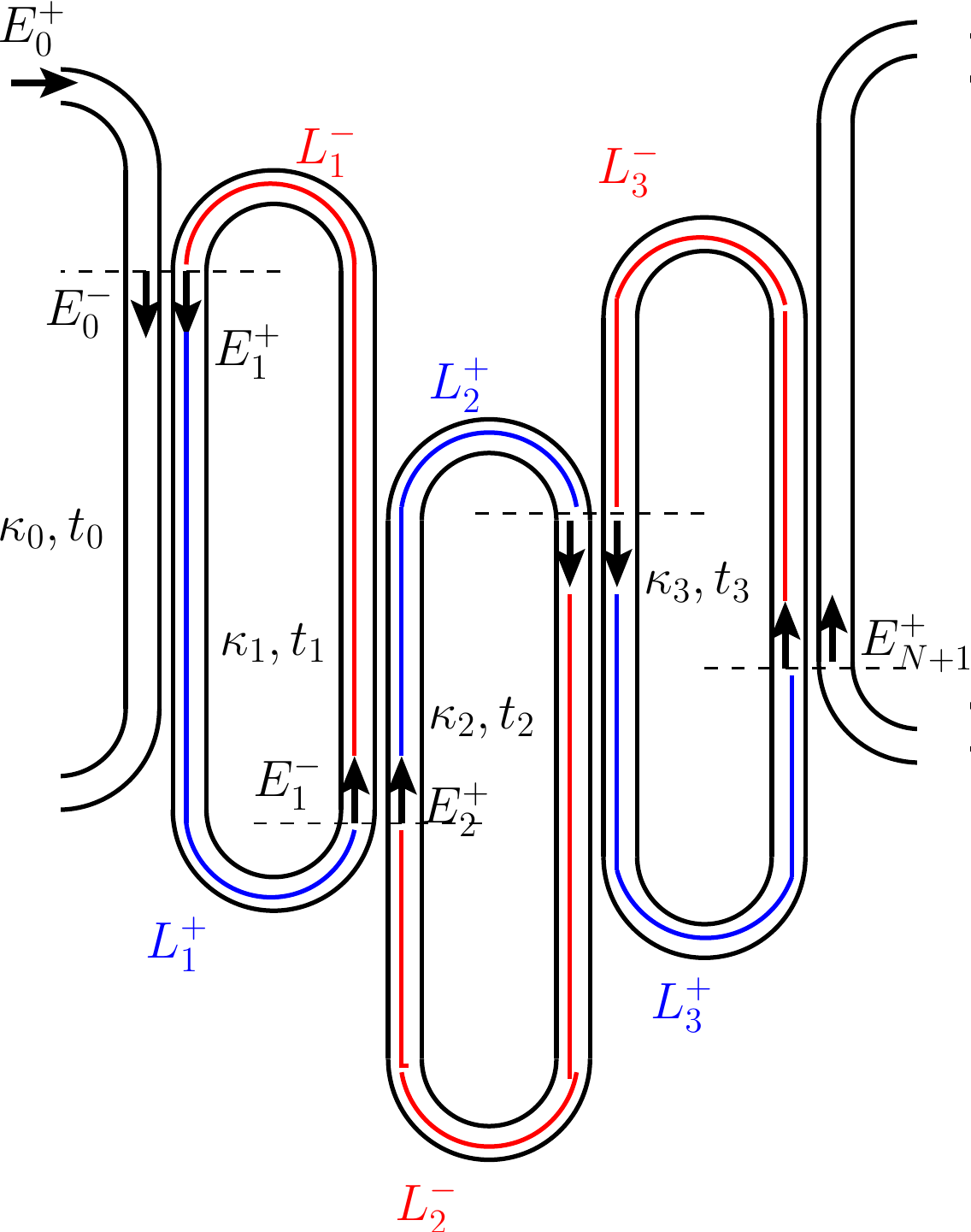}
\end{tabular}
\caption{Layout of a $N=3$ ring CROW with longitudinal offset coupling. (a) with all the lateral displacements produced downwards, (b) with alternating directions for the lateral displacements and (c) three rings with the two first couplings arranged downwards and the two last ones upwards, as in Ref \cite{domenech2009}. }
\label{esquema}
\end{figure}

Coupled microring structures are small-size, compact optical devices which provide
photonic processing functionalities such as optical filtering
\cite{Lenz,jlt}, dispersion compensation \cite{Madsen}, optical sensing
\cite{Chao}, or the control of the group delay (slow and fast light)
\cite{Boyd}, \cite{Pedro1}. Group delay control is particularly important for
the development of high-performance photonic networks, where ways to obtain
large, distorsionless propagation delays are currently sought after. The
research in slow light has made intensive use, at least as a starting point,
of the two \textquotedblleft basic\textquotedblright\ MRR structures : the
Side Coupled Integrated Spaced Sequence of Resonators (SCISSOR) \cite{Heebner}%
, and the Coupled Resonator Optical Waveguides (CROW) \cite{Poon}. Unlike the
SCISSOR, which is an all-pass structure, that is, a pure phase filter, the
CROW has four ports and is functionally equivalent to a Bragg grating.
Motivated by this analogy, the technique of apodization or windowing, borrowed
from the field of digital filter design and regularly employed in grating
filters \cite{Cap96}, has also been proposed to improve the characteristics of
CROW structures \cite{Cap2007}. Basically, apodizing the coupling constants of
the coupled waveguides in the unit cells helps to reduce the side-lobes in the
SCISSOR and the ripples in the pass-band of the CROW.

A general difficulty of the MRR designs is to attain the high accuracy
required in the values of the coupling constants ($K$) when it comes to the
fabrication process. Typically, the adjustment of the $K$ values is
acomplished by changing the separation between the waveguides of the coupler
(transversal offset). This is a demanding technique which restricts the
possibilities of design because it may require, along the longitudinal
direction, resolution steps below the the current fabrication limit (of the
order of a few nm). To overcome this difficulty, a much less stringent
technique was proposed in \cite{domenech2009}. Here the coupling constant
values are adjusted by longitudinal offset between the coupled waveguides,
while the inter-guide separation is left fixed. The resolution requirements
are relaxed by two orders of magnitude, making photolitographic production
feasible. This technique has been demonstrated in \cite{domenech2011} for a
3-racetrack apodized CROW structure fabricated in silicon-on-insulator.

Despite the general good agreement between the expected and the measured
results in the structure considered in \cite{domenech2011}, we identify in this
letter a fundamental problem which is inherent to all apodized CROW structures
designed by the longitudinal offset technique. We will review the derivation
of the CROW transfer function, paying particular attention to the
role of the directional couplers implemented using the longitudinal displacement method. 
With the obtained results, we
demonstrate the actual performance degradation occurring in these devices.

\section{Model}

The geometry of a three ring CROW with longitudinal displacement couplings is shown in Fig. \ref{esquema}, including the definition of the field variables $E_l^+$ and $E_{l-1}^-$. $E_{N+1}^-$ will be assumed to be null and its phase reference is left undefined.  $L_c$ is the length of the straight waveguide sections and corresponds to the coupler length for zero lateral desplacement, $L_R=\pi R$,  is the length of each curved waveguide section of radious $R$. The total ring length is, therefore, $L_T=2L_R+2L_C$ and will be assumed constant in this work.  For simplicity, we will ignore the actual correction in the coupling length due to the curved sections \cite{xia} and set $L_{eff}=L_c$.

It is well known that the acummulated propagation phase for
any of the output field components of the lossless coupler of length $L$
is $\exp(j\beta L)$ \cite{yariv}, where $\beta$ is the propagation constant of
the mode of each of the individual waveguide sections that are combined in the
coupler. Therefore, if we call $E_{o,1}$ $E_{o,2}$ the complex field
amplitudes at the output ports of a coupler, and $E_{i,1}$ $E_{i,2}$ those at
the input, we have
\begin{equation}
\left [
\begin{array}{c}
E_{o,1}\\
E_{o,2}
\end{array}
\right ]=
\left [ 
\begin{array}{cc}
t & \kappa\\
-\kappa^*&t
\end{array}
\right ]
\left [
\begin{array}{c}
E_{i,1}\\
E_{i,2}
\end{array}
\right ]\exp\left(j\beta L\right) \label{coupler}
\end{equation}
with $\kappa$ and $t$ the parameters defining the coupler behavior.

The common phase factor $\exp(j\beta L)$ permits to model the coupler as a lumped element localized at the
 input plane and described by Eq. \eqref{coupler} excluding the phase term that is included in the propagation model.  The positions of the coupling points are marked with horizontal dashed lines in Figure \ref{esquema} and correspond to the reference planes for the field variables.

The equation relating the fields in the rings $l$ and $l+1$
can be readily written with a general notation as%
\begin{equation}
\left[
\begin{array}
[c]{c}%
E_{l+1}^{+}\\
E_{l+1}^{-}%
\end{array}
\right]  =\mathbf{M}_{l}\left[
\begin{array}
[c]{c}%
E_{l}^{+}\\
E_{l}^{-}%
\end{array}
\right]
\end{equation}
with
\begin{equation}
\mathbf{M}_{l}=\dfrac{1}{\kappa_{l}}\left[
\begin{array}
[c]{cc}%
-\left(  |\kappa_{l}|^{2}+|t_{l}|^{2}\right)  \exp(j\delta_{l}^{+}) &
t_{l}^{\ast}\\
-t_{l}\exp\left[  j\left(  \delta_{l}^{+}-\delta_{l+1}^{-}\right)  \right]   &
\exp(-j\delta_{l+1}^{-})
\end{array}
\right]
\end{equation}
and $\delta_{l}^{+}=\beta L_{l}^{+}$, $\delta_{l}^{-}=\beta L_{l}^{-}$,
where we have assigned a total propagation length $L_{l}^{+}$ 
to the forward propagating signal in ring $l$,  and a length
$L_{l}^{-}=L_{T}-L_{l}^{+}$  to the backward component, as
illustrated in Figure \ref{esquema}. For simplicity, we have assumed
negligible propagation losses.  The expressions of $L_{l}^{+}$ 
and $L_{l}^{-}$ are given below for some specific cases.

For an ideal evanescent coupler of the type described in Figure
\ref{esquema}, we have $\kappa_{l}=j\sqrt{K_{l}}$ and $t_{l}%
=\sqrt{1-K_{l}}$, where $K_{l}$ is the coupling constant for
the $l$-th coupler. The overall output-input relation can then be
written as
\begin{equation}
\left[
\begin{array}
[c]{c}%
E_{N+1}^{+}\\
E_{N+1}^{-}%
\end{array}
\right]  =\mathbf{M}\left[
\begin{array}
[c]{c}%
E_{0}^{+}\\
E_{0}^{-}%
\end{array}
\right]
\end{equation}
with
\begin{equation}%
\begin{split}
\mathbf{M} &  =(-j)^{N}\prod_{l=0}^{N}\dfrac{1}{\sqrt{K_{l}}}\times\\
&  \left[
\begin{array}
[c]{cc}%
-\exp(j\delta_{l}^{+}) & \sqrt{1-K_{l}}\\
-\sqrt{1-K_{l}}\exp\left[  j\left(  \delta_{l}^{+}-\delta_{l+1}^{-}\right)
\right]   & \exp(-j\delta_{l+1}^{-})
\end{array}
\right]  .
\end{split}\label{matriz}
\end{equation}

The exact expressions of the propagation phase shifts will vary
for each particular geometry implemented.  For an all-downwards arrangement
as the one depicted in Figure \ref{esquema} (a) we have, for $l\neq0$ and $l\neq
N,$%
\begin{equation}
L_{l}^{+}=\left\{
\begin{array}
[c]{ll}%
L_{C}+L_{R}+L_{\text{off}}^{l}-L_{\text{off}}^{l-1} & \qquad\text{if }l\text{
even}\\
L_{C}+L_{R} & \qquad\text{if }l\text{ odd,}%
\end{array}
\right.\label{down}
\end{equation}
and $L_{l}^{-}=L_{T}-L_{l}^{+}$.

For an alternating arrangement, such as the one shown in Figure \ref{esquema}%
.b, we have for $l=1,\dots,N-1$,
\begin{equation}
L_{l}^{+}=L_{c}+L_{R}+L_{off}^{l-1}+L_{off}^{l},
\end{equation}
and $L_{l}^{-}=L_{T}-L_{l}^{+}$. 

In either case we have $\delta_{0}^{+} =\beta\left(L_{R}/2+L_{off}^0\right)$ if $l=0$  and, if $l=N$, $\delta_{N+1}^{-}=0$.

\section{Results and discussion}

 In general, the phase asymmetry will greatly affect the system response
in such a way that any resemblance with the original target transfer function can be 
completely lost, but the effect is largely dependent of the filter geometry
and can be controlled if it is considered in the design stage of the optical filter.  

If the all-downwards
configuration of Figure \ref{esquema} (a) is chosen in the implementation of a CROW 
with constant coupling coefficients using the longitudimal 
displacement technique,  there is no 
phase asymmetry since $L_{off}^{l}$ is constant and the terms producing the asymmetry 
in \eqref{down} cancel out.  Similarly, it can be shown that the amplitude response obtained in the
non-apodized case with the alternating configuration of Figure \ref{esquema} (b) 
corresponds to that of the ideal case.

\begin{figure}[!t]
\centering
\includegraphics[width=3.5in]{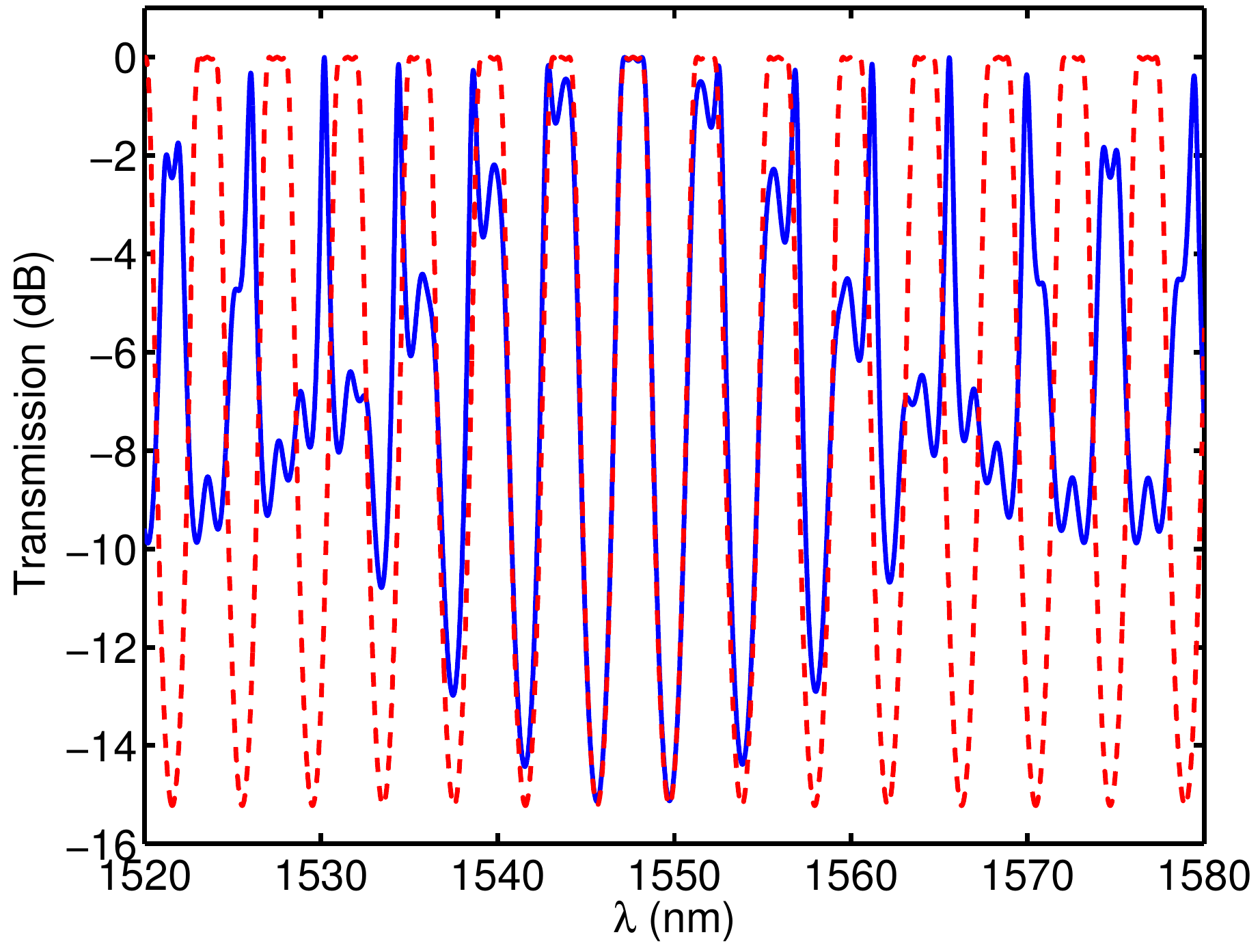}
\caption{Transmission spectra for the three-ring CROW design  with (solid line) and without (dashed line) the effect of phase asymmetry.}
\label{espectro}
\end{figure}

However, in apodized filters the phase asymmetry produced by the variation of the offset 
lengths along the CROW waveguide introduced to modulate the coupling coefficients results
in a severe distortion of the system transfer function as compared to the targed design. 

We apply our findings to the experimental results reported in \cite{domenech2011}. 
We use the same parameters as in the $N=3$ fabricated filter with $n_g=2.24$, $R=\SI{5}{\micro\meter}$, $L_s=\SI{53.3}{\micro\meter}$ coupling constants $\left\{K_l\right\}_{l=0}^3=\left\{0.86, 0.54, 0.54, 0.86\right\}$ and $\left\{L_{off}^l\right\}=\left\{\SI{10.56}{\micro\meter}, \SI{2.42}{\micro\meter} ,\SI{2.42}{\micro\meter}, \SI{10.56}{\micro\meter} \right\}$.  The geometry is that of Figure \ref{esquema} (c).  In this case we have  $L_1^-=L_c+L_R$, $L_2^-=Lc+L_R+L_{off}^1$, $L_3^-=L_C+L_R+L_{off}^2-L_{off}^3$ and $L_l^+=L_T-l_l^+$.  The amplitude response of the ideal apodized filter is shown with dashed lines in Figure \ref{espectro}, whereas the calculated amplitude response including the phase asymmetry is shown in the same figure with solid lines.   

In spite of our simplifying assumptions, the correspondence of the calculated curve with the experimental measurements in the spectral region beween $\SI{1550}{\nano\meter}$ and $\SI{1554}{\nano\meter}$ displayed in the Figure 4 of Reference \cite{domenech2011} is remarkable. 

The effect of the phase asymmetry is shown to destroy the periodicity of the system response.  Even though a very good fit with the ideal response is found for a limited spectral range, the resemblance with the target system response can be completely lost in other spectral regions.  We note that in this case it is by chance that good correspondence has been obtained in the region of interest, since no a priori provision for the effect of phase asymmetry was made in the design.

In conclusion, we have highlighted the existence of a phase asymmetry effect in the implementation of CROW filters using the lateral displacement technique.  We have shown that even though certain filter geometries of non-apodized filters can be safely implemented, this effect cannot be ignored in the design of apodized CROWS.

\end{document}